\documentclass[conference]{IEEEtran}
\usepackage{cite}

\usepackage[pdftex]{graphicx}
\graphicspath{{./Figures/}}
\DeclareGraphicsExtensions{.pdf,.jpeg,.png}
\usepackage{array}
\usepackage{url}

\usepackage[acronym, nonumberlist, nogroupskip, nopostdot]{glossaries}
\usepackage[absolute]{textpos}
\usepackage{soul}
\newcommand{\nhl}{}

\newglossaryentry{AS}{
    name=Ancillary Service,
    description={An ancillary service describes a service provided by one or more resources to the power system in order to maintain is stability and power quality.}
}

\newacronym{DER}{DER}{distributed energy resource}
\newacronym{TSO}{TSO}{transmission system operator}
\newacronym{DSO}{DSO}{distribution system operator}
\newacronym{PV}{PV}{photovoltaics}
\newacronym{DRES}{DRES}{distributed renewable energy source}
\newacronym{ICT}{ICT}{information and communication technology}

\makeglossaries

\begin{document}

\title{A Tutorial on Resilience in Smart Grids}

\author{\IEEEauthorblockN{Armin Stocker}
\IEEEauthorblockA{Chair of Computer Networks and\\Computer Communication\\
University of Passau\\
Passau, Germany 94032\\
Email: Armin.Stocker@uni-passau.de}
\and
\IEEEauthorblockN{Hermann de Meer}
\IEEEauthorblockA{Chair of Computer Networks and\\Computer Communication\\
University of Passau\\
Passau, Germany 94032\\
Email: Hermann.DeMeer@uni-passau.de
}
}


%

\IEEEspecialpapernotice{2022 International Workshop on Resilient Networks Design and Modeling (RNDM)}
\begin{textblock}{18.5}(1.5, 0.25)
\noindent © 2022 IEEE. Personal use of this material is permitted. Permission from IEEE must be obtained for all other uses, in any current or future media, including reprinting/republishing this material for advertising or promotional purposes, creating new collective works, for resale or redistribution to servers or lists, or reuse of any copyrighted component of this work in other works. \newline
DOI: 10.1109/RNDM55901.2022.9927711
\end{textblock}

\maketitle

\begin{abstract}

A key quality of any kind of system is its ability to deliver its respective service correctly.
Often the unavailability of commercial systems may lead to lost revenue, which are minor compared to what may be at stake when critical infrastructures fail.
A failure to deliver critical services, such as clean water or electricity may have dire consequences that endanger human lives and may even halt or break other infrastructures.
The services provided by critical infrastructures need to be supplied continuously even when faced with re-configurations, outside disturbances and systemic changes.
A system is called resilient if it fulfils this property.
From the many critical infrastructures that exist, power systems may be the most important ones, because they are supplying the required electricity for other critical infrastructures.
At the same time, a power system itself may be exposed to several disturbances from internal sources, e.g., fluctuations in the energy demand, and external sources, e.g., heavy storms.
Especially, fast dynamic effects caused by these disturbances may lead to deviations of grid frequency, short-circuits, or, in severe cases, a total power system failure.
As future scenarios will include more distributed renewable sources and less centralized generation from fossil fuels, ICT-based communication and coordination will play an increasing role.
This paper examines the notion of resilience, how it has been traditionally ensured for the power system, and novel approaches to maintain the frequency, protect people and devices against short circuits and recover from a blackout.
A special focus is on communication and the role that distributed renewable generation plays for these processes.


\end{abstract}


%
\IEEEoverridecommandlockouts
\IEEEpubid{\makebox[\columnwidth]{978-1-5386-5541-2/18/\$31.00~\copyright2022 IEEE \hfill} \hspace{\columnsep}\makebox[\columnwidth]{ }}
\IEEEpeerreviewmaketitle
\IEEEpubidadjcol

\section{Introduction}\label{sec:Introduction}

To reduce carbon emissions of the power supply and to reduce the reliance on fossil fuels, an increasing amount of renewable energy sources are connected to power grids.
At the same time, large, fossil fuel or nuclear generators are decommissioned.
Apart from supplying sufficient energy to match the demand for electricity, however, the operation of the power system requires several other services to remain stable. These services are known as ancillary services.
Many of the ancillary services have been provided inherently by the physical characteristics of rotating masses and electromagnetic fields of synchronous generators as an integral constituent of fossil fuel and nuclear generators.
Therefore, a decommissioning of these synchronous generators leads to a reduction in the availability of ancillary services.
Based on these observations, the ENTSO-E \cite{entsoe_inertia} predicts, for example, insufficient resources for inertial reserves in many European countries by 2030. Inertia is one of the most important system characteristics for the power system as it immediately opposes fluctuations in the grid caused by momentary imbalances between demand and supply.
The inertial reserve is quantified as the number of seconds for which energy reserves to oppose a large frequency deviation are available.
This is also referred to as the inertia constant commonly denoted as $\text{H}$.
Traditionally, about five to six seconds of inertial reserves were available and were considered as a sufficient safeguard\cite{entsoe_inertia}.
As can be seen in Table~\ref{tab:inertiaByCountry}, inertial reserves larger than four seconds, as the absolute lower bound, are predicted to be available only in a few remaining countries, whereas in most European countries only a limited or marginal amount of inertia, which is considered as insufficient and unsafe, will be provided.
Inertial response is the most prominent ancillary service required for a stable operation of the power grid.
A lower system inertia leads to a lower tolerance of major disruptive events, such as a large generator failing.
As a result, the dependability of a power system is decreasing due to these changes.
Maintaining dependability despite large changes is at the core of resilience.
Therefore, a resilient power system requires alternative ways to provide ancillary services.


\begin{table}
    \centering
    \caption{Inertia constant prediction for 2030 by the ENTSO-E \cite{entsoe_inertia}.}
    \begin{tabular}{|c|c|c|} \hline
    Country         & Qualitative & Quantitative \\
                    & Contribution & Contribution  \\ \hline
    Belgium         & Limited   & $\text{H} \leq
    2$s \\
    Croatia         & Limited   & $\text{H} \leq
    2$s \\
    Germany         & Limited   & $\text{H} \leq 2$s \\
    Greece          & Limited   & $\text{H} \leq
    2$s \\
    Ireland         & Limited   & $\text{H} \leq
    2$s \\
    Italy           & Limited   & $\text{H} \leq 2$s \\
    Luxembourg      & Limited   & $\text{H} \leq
    2$s \\
    Portugal        & Limited   & $\text{H} \leq 2$s \\
    Spain           & Limited   & $\text{H} \leq 2$s \\
    United Kingdom  & Limited   & $\text{H} \leq
    2$s \\\hline
    
    Austria         & Marginal & $2$s$\leq \text{H} \leq 3$s \\
    Albania         & Marginal & $2$s$\leq \text{H} \leq 3$s \\
    Bulgaria        & Marginal & $2$s$\leq \text{H} \leq 3$s \\
    Denmark         & Marginal & $2$s$\leq \text{H} \leq 3$s \\
    Netherlands     & Marginal & $2$s$\leq \text{H} \leq 3$s \\
    Switzerland     & Marginal & $2$s$\leq \text{H} \leq 3$s \\\hline
    
    Bosnia and Herzegovina         & Good      & $3$s$\leq
    \text{H} \leq 4$s \\
    Finland         & Good      & $3$s$\leq \text{H} \leq 4$s \\
    France          & Good      & $3$s$\leq \text{H} \leq 4$s \\
    Latvia          & Good      & $3$s$\leq \text{H} \leq 4$s \\
    Norway          & Good      & $3$s$\leq \text{H} \leq 4$s \\
    Romania         & Good      & $3$s$\leq \text{H} \leq 4$s \\
    Sweden          & Good      & $3$s$\leq \text{H} \leq 4$s \\\hline
    
    Estonia         & Very Good & $4$s$\leq \text{H}$ \\
    Hungary         & Very Good & $4$s$\leq \text{H}$ \\
    Montenegro      & Very Good & $4$s$\leq \text{H}$ \\
    Poland          & Very Good & $4$s$\leq \text{H}$ \\
    Slovakia        & Very Good & $4$s$\leq \text{H}$ \\
    Serbia          & Very Good & $4$s$\leq \text{H}$ \\\hline
    \end{tabular}
    \label{tab:inertiaByCountry}
\end{table}



If sufficient ancillary service provision cannot be ensured, a lower fault tolerance and lower system stability proportionally increases the risk of blackouts.
One way to reverse this trend is for \glspl{DER} to become primary providers of both energy and ancillary services.
From a power systems standpoint, this may be achieved with smart converters coupled with renewable generation sources \cite{EasyRESInertia}, \cite{ieee9508455}, an introduction of storage systems, such as community scale batteries or flywheels, or by exploiting demand-side flexibility, e.g., by enabling smart charging strategies for electrical vehicles as suggested in \cite{ieee6268312}.
Integration of renewable energy sources on a large scale has other effects that invalidate assumptions made regarding a top-down power flow. As an example for this, the protection against short circuits or other grid faults must be adapted to the new conditions.
To ensure sufficient ancillary service reserves are available before and during a disruption of the normal grid operation, an \gls{ICT} system coordinating these resources is required.



Smart grid refers to this type of interconnected ICT and power system.
As a power system is a critical infrastructure, the requirement to ensure a dependable operation even when unforeseen circumstances arise extends to ICT components of a smart grid.
Therefore, any smart grid needs to be built with an understanding of the requirements for resilient operation.

The characteristics of renewable sources are both a challenge and an opportunity.
As the electricity provided by wind farms or \gls{PV}-plants depends on weather conditions, it is predictable only to a certain extent.
Additionally, distributed generation sources are generally of lower capacity and designed with lower reliability standards than carefully engineered synchronous generators.
Taken on its own, each of these devices is therefore more likely to fail.
These two considerations increase the internal challenges that a power system needs to overcome while at the same time exhibiting the same tolerance of any external disturbances as before.
However, by means of ICT-based coordinated operation of these new \gls{DER}, an increasing number of generation sites and the increasing location diversity can be leveraged to maintain the system stability and recover from disturbances.



This paper highlights the new obstacles for frequency control, the short-circuit protection and blackout recovery. For each of these, approaches from literature that make the power system dependable despite these changes are reviewed and the role that an resilient ICT system has to fulfill is discussed.
To do so, the rest of this paper is structured as follows:
First, the resilience concept is introduced in Section~\ref{sec:Background}. 
Afterward, frequency maintenance, short circuit protection and blackout recovery are discussed in Section~\ref{sec:Frequency}, Section~\ref{sec:ShortCircuit}, and Section~\ref{sec:Blackout}, respectively.
Finally, Section~\ref{sec:Conclusion} concludes the paper.

\section{Background on Resilience}\label{sec:Background}


Resilience definitions are slightly different in literature.
The ResiliNets project \cite{sterbenz:2010:RAS} defines resilience as the ability of the network to provide and maintain an acceptable level of service in the face of faults and challenges.
According to \cite{ieee8473549} resilience is the ability of a system to continuously provide its critical functionality and rapidly recover from cyber-attacks.
\cite{ieee9281596} measures resilience using the recovery rate, number of survivable, coinciding failures and duration of events.
A further definition of resilience is given in \cite{ieee9526051} as the ability of a network to bear threats, absorb disturbances, operate at low power, and restore its functional state within an acceptable time.
\cite{ieee8974395} stresses the importance to focus on high impact low probability events and classifies measures for power system resilience into planning, response and restoration in its definition of resilience.
\cite{ieee7819910} notes that resilience is the ability of a system to recover rapidly after internal and external disruptions.

While these references put a greater focus on cyber-attacks or exceptional threats, overall these definitions are similar to the definition by \cite{Laprie_Resilience} as adopted in this paper: Resilience is ``the persistence of dependability when facing changes".
A system is considered dependable if one can place justifiable trust in the delivery of its service.
That means dependability requires availability, reliability, performance, maintainability, safety and security.

This paper chooses to focus on resilience in the context of smart grids, because power systems currently face major changes.
According to \cite{Laprie_Resilience} a change can be classified based on its nature (functional, environmental, or technological), its duration (seconds to hours, hours to months, or months to years) and finally its predictability (foreseen, foreseeable, or unforeseen). 
Examples of these changes to the power system and their classification are the following:
\begin{enumerate}
    \item A foreseen, long-term, technological change to the power system is caused by the massive increase in renewable generation that is replacing fossil fuel plants and the corresponding reduction of ancillary service reserves described in Section~\ref{sec:Introduction}.
    \item A foreseeable, long-term, functional change may be caused by an increase in electric vehicles and heat pumps over the coming years. This changes the demand profile of households drastically, because the energy used for heating and mobility then is to be supplied by the power system. 
    \item  Drastic events, such as the recent pandemic with widespread lockdowns or the war in Ukraine that lead to a shortage of fossil fuels, lead to unforeseen, long-term, environmental changes.
    \item A change that leads to short-term, foreseeable, functional effects is the reversal of the power flow caused by an increasing production of renewable sources in a distribution network. A reversed power flow requires adapting the protection system to this new condition to maintain safety.
\end{enumerate}


The changes in these examples impact the availability of ancillary services, the expected performance of the system, the operation of the safety mechanisms or, in short, the dependability of the power system. Therefore, mechanisms are required to restore dependability and, as a result, make the power system resilient.

\subsection{Resilience Framework}

This section introduces the framework used in this paper to structure the description of existing resilient systems.
These notions will be used to explain the processes occurring in future smart grid scenarios with special relevance to an accompanying \gls{ICT} systems.

\subsubsection{$\text{D}^2\text{R}^2+\text{DR}$}

\begin{figure}
    \centering
    \includegraphics[width=0.70\linewidth]{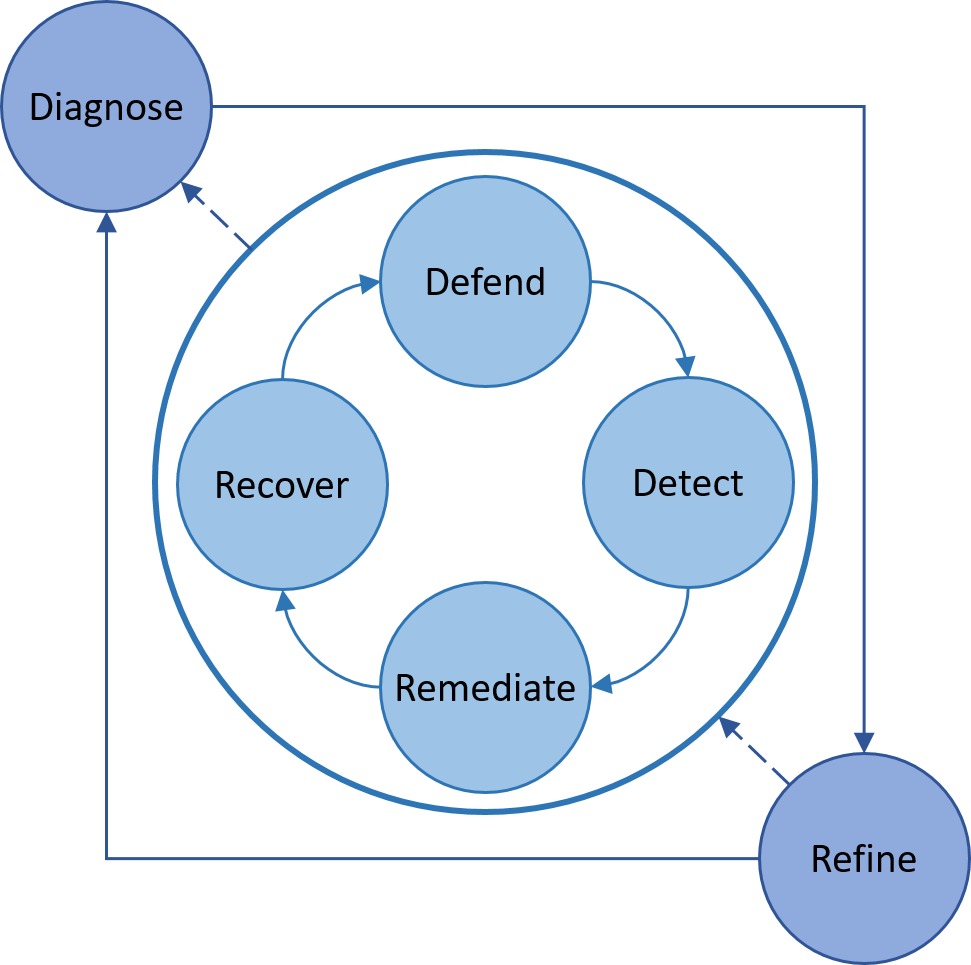}
    \caption{$\text{D}^2\text{R}^2+\text{DR}$ resilience framework (adopted from ResiliNets\cite{sterbenz:2010:RAS}).}
    \label{fig:D2R2}
\end{figure}

The framework proposed within the ResiliNets project is used to classify the operational stages of a resilient system. This framework is shown in Fig.~\ref{fig:D2R2}.
An inner control loop is formed by four concepts: ``Defend", ``Detect", ``Remediate", ``Recover". While an outer control loop for the continuous learning and improvement of the system is created by the concepts ``Diagnose" and ``Refine". 

``Defend" summarizes the measures put in place to maintain the system in its original operational state. An example of such a measure may be a firewall blocking externals from accessing internal resources of the system.

``Detect" aims to find disturbances to a systems operation and initiate the proper reaction of the system. For example, scanning traffic for suspicious patterns as part of an intrusion detection system can detect an ongoing cyber-attack.

``Remediate" are intermediary measures minimizing the impact of a challenge in the short term.
These measures increase the level of service during a disturbance by better utilizing a deteriorated system state or partially restoration of the state. An example is the upscaling of existing virtual machines while a denial-of-service attack or a flash crowd is happening to maintain service provision to legitimate customers.

The final phase of the inner loop, ``Recover", removes the cause for the disturbance restoring the original system state and level of service. Refining the firewall rules to block the illegitimate traffic of the denial-of-service attack is one way to end the ongoing attack.

Compared to the term ``challenge" introduced in \cite{sterbenz:2010:RAS}, a change is a drastic alteration of the systems state or environment. This change can then lead to a different set of challenges to the normal operation of a system that must be overcome to maintain the systems service.
The ``Diagnose" and ``Refine" stages enable adaptation to these changes.

After a change, ``Diagnose" aims to find the root causes to be considered for challenges as a basis for the improvement of the system. Finally, ``Refine" is the increase of system resilience by creating new measures and fine-tuning existing mechanisms to better withstand future disturbances. The role of the outer loop is to enable the system to adapt itself. An ideal resilient system is able to adapt its behavior when facing systemic changes.

\subsubsection{Resilience State Space}
\begin{figure}
    \centering
    \includegraphics[width=0.9\linewidth]{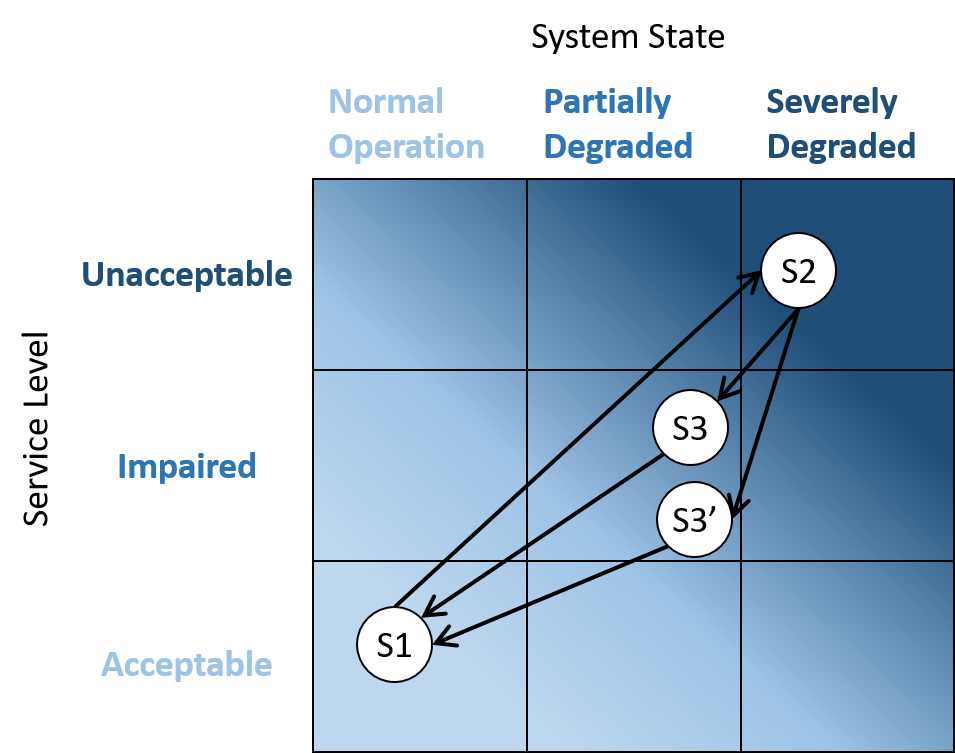}
    \caption{Resilience state space.}
    \label{fig:ResilienceState}
\end{figure}

The resilience state space introduced in \cite{SystemicResilience} is a way to depict resilience as a relation between degradation of the system state and the available level of service. 
A 2-dimensional space is partitioned into a service level (vertical) and a system state (horizontal) axis.
Therefore, to place any given system configuration in this 2-d space, the operational system state and the available service level for that system configuration needs to be determined.

State transitions caused by a challenge or restoration action are indicated as arrows between states showing the impact on system state and service level.
Thereby, the relation of challenges, system state and restoration can be illustrated.
A simplified figure showing a possible case for the service degradation after an outage is shown in Fig.~\ref{fig:ResilienceState}.

It is noteworthy, that there indeed is a relation between the system state and the service level. However, taking this to be a 1-to-1 mapping is wrong. As an example, a degraded communication system where routers have failed may perform worse initial until new routes can be computed. The operational condition remains degraded, yet these new routes may re-establish communication that previously involved one of the failed nodes.

Differences in the resilience of a system are noticeable in this depiction by considering the trajectory of arrows in the space.
An arrow depicting a degradation of a system, e.g., from S1 into S2, will have less area underneath it if the system has a higher resilience.
Take for example a system whose components are better suited to survive a challenge or a system that is better at dealing with failed components compared to the one in Fig.~\ref{fig:ResilienceState}. This system will have shorter arrows for the transition after a challenge, because the system state and service level are degraded less.
As a result, for arrows representing a state degradation the area underneath will decrease if resilience increases.
Similarly, if remediation is improved the area under the arrow from S2 to S3 will be larger, because the remediation has a bigger impact on the service level provided by the system. Such an improvement is illustrated as the remediation from S2 to the state S3'.

\begin{figure}
    \centering
    \includegraphics[width=\linewidth]{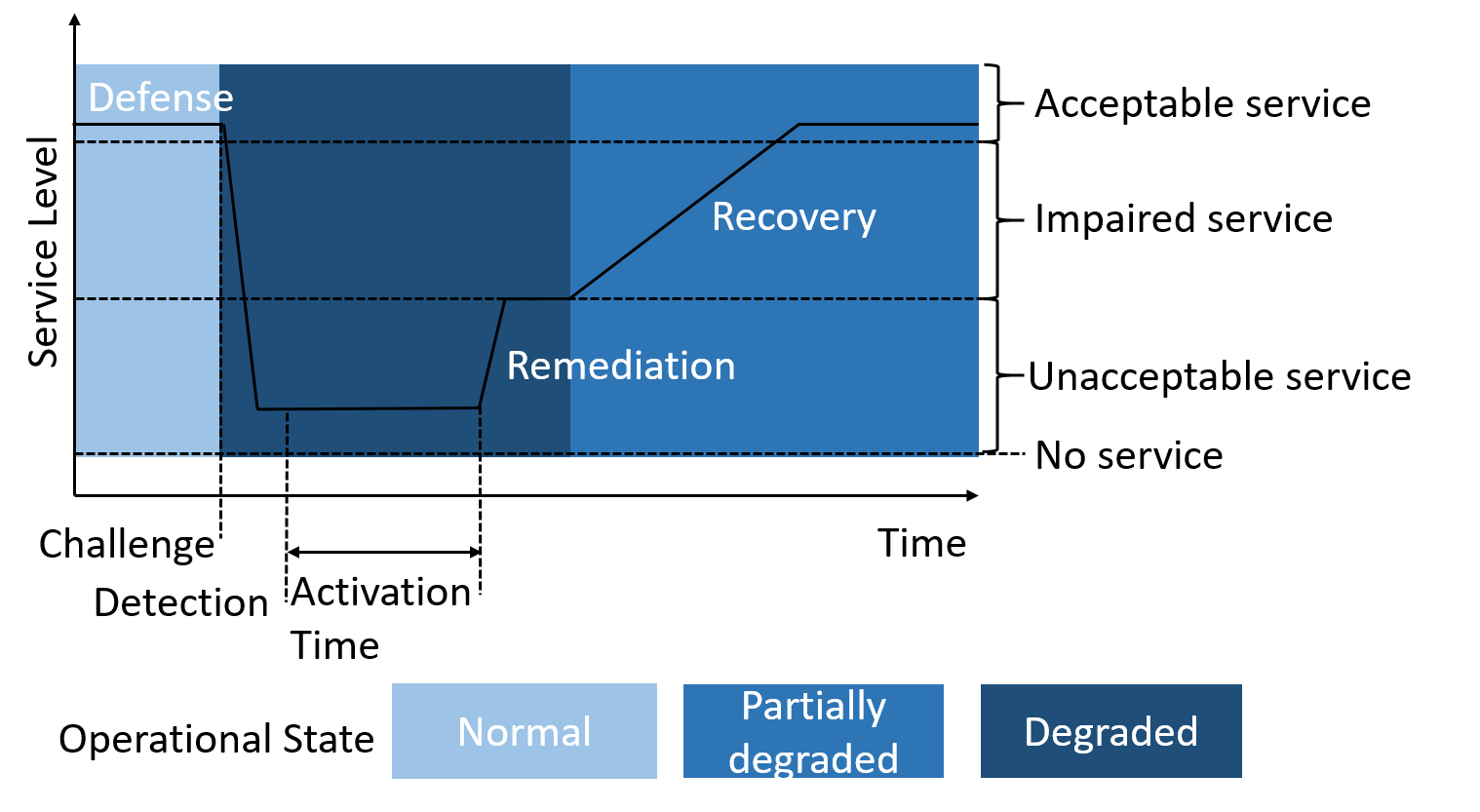}
    \caption{Service level of a resilient system over time.}
    \label{fig:ResilienceOverTime}
\end{figure}

Insights into the behavior of a system can also be gained by considering the time dimension of a disturbance as shown in Fig.~\ref{fig:ResilienceOverTime}.
The curve follows the same four phases of the inner control loop as depicted in the resilience state space.
At the start, the system is kept in its original operational state by ``Defense" measures. 
After some time, an unexpected challenge occurs degrading the operational state and as a consequence the service level.
The detection mechanism will trigger sometime after the adverse event.
Afterward, the process of selecting and activating the appropriate measures for ``Remediation" begins.
These restore the systems service level to a less degraded state, which lasts until the recovery mechanism can restore the original service level.
This depiction additionally represents the time between challenge and detection, the activation time between detection and remediation, the time for remediation and the time for recovery.

Therefore, an improved system resilience after ``Diagnose" and ``Refine" can also be achieved by one of the following.
First, adding defensive measures can avoid system deterioration for a longer time.
Second, faster acting detection and a shorter activation time until remediation starts, decreases the time at the lowest service level.
Finally, improving remediation and recovery can lead to these processes finishing faster and restoring the system to a state that provides a higher service level despite the degraded operational state.

With the basics of resilience discussed here, the selection of ancillary services presented in the remainder of the paper can be better understood as well.
Frequency and inertial reserves are what enables a power system to survive and recover from the failure of large generation sources. Therefore, they are a corner stone to a power systems dependability that needs to be investigated.
The shift from centralized generation to distributed resources also impacts the operation of the protection system that is an important safety mechanism inside a power grid.
Therefore, a dependable power system requires maintaining the proper operation of this sub-system as well.
Finally, blackouts are the most severe condition that can occur in a power system and lead to the most challenging recovery situations. 
This is especially true in scenarios in which this recovery needs to happen bottom up due to an upstream grid with only a few small synchronous generators.  
These three ancillary services are identified as being impacted the most by the changes to power systems outlined in Section~\ref{sec:Introduction} and are selected to be discussed in this paper.
\section{Frequency Maintenance}\label{sec:Frequency}


The current power system in Europe operates at a nominal frequency of 50 Hertz.
For the stable operation of the power grid, it is important that this parameter is controlled to be within +-0.5 Hz.
At its core, the frequency of the grid is determined by the instantaneous balance of power supply and consumption.
As the connected loads are constantly changing in small quantities, the frequency in the grid is also exposed to small disturbances.
However, larger disturbances, which may be created by an industrial scale electric motor being started or a large generator failing, require ancillary services to maintain stability and eventually restore the normal service level provided by the system.

\begin{figure}
    \centering
    \includegraphics[width=\linewidth]{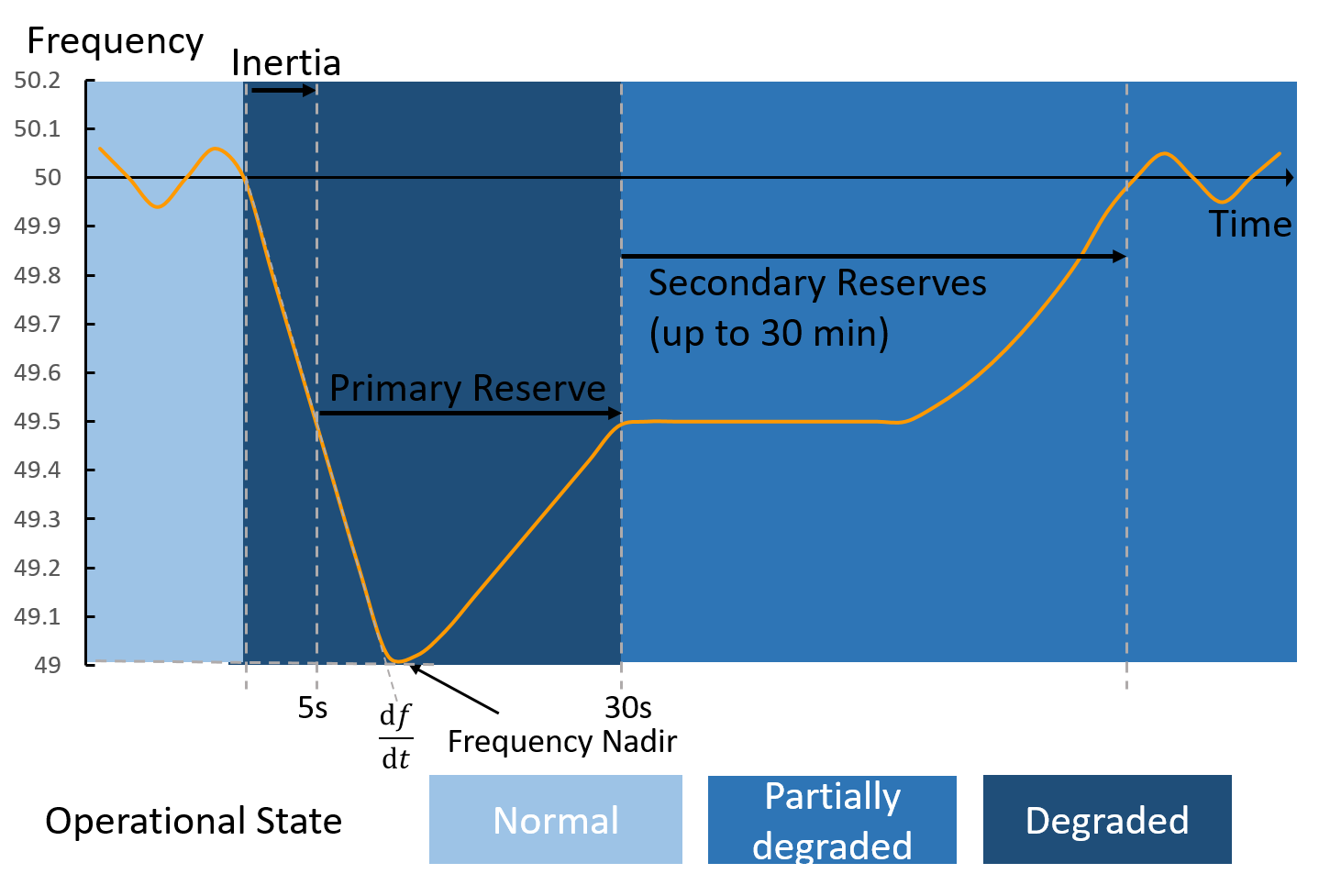}
    \caption{Reaction to a large frequency disturbance.}
    \label{fig:FrequencyOverTime}
\end{figure}

The frequency development caused by a disturbance and the response of the system to this is shown in Fig.~\ref{fig:FrequencyOverTime}. \nhl{This behavior aligns well with the detection, remediation and recovery phases of the ResiliNets Framework. This mapping is elaborated in the following.}
\nhl{The service level is determined based on the frequency of the power system and the operational state for this use case is depending on the state of the generators in the system.}
As soon as the frequency changes drastically, the inertia of rotating masses of synchronous machines counteracts this change.
This dampens the rate of change of frequency $\frac{\text{d}f}{\text{d}t}$ and increased the frequency nadir, i.e., the minimum value the frequency drops to, is increased. 
\nhl{This drop needs to be slowed down such that the detection mechanism and activation period of the primary frequency reserves can happen. Only then can these reserves start the remediation of the frequency.} As stated previously, traditionally inertia reserves were sized to cover the first 5 seconds of an incident.

In the European grid, the primary frequency reserves are tendered as so-called frequency containment reserves.
The frequency containment reserve is provided by a set of flexible resources that are able to change their power output with certain ramping requirements.
The resources need to be able to ramp to 50\% of the maximum output after 15 seconds and the full output after 30 seconds.
\nhl{The tradeoff for this fast response is that the resources are insufficient to recover the frequency completely.}
After the primary frequency reserves, secondary frequency reserves are activated to recover the frequency to its nominal value over a longer period.
The secondary frequency reserves take a longer time to reach their maximum output, but are required to sustain this supply for longer.
During the provision of secondary frequency reserves, also the spent primary frequency reserves are recovered.
\nhl{As these resources are ramping up and connecting to the grid, an argument can be made as to an improvement of the grids operational state from severely degraded to degraded because new generation resources are added. However, this state is still worse than the original state as the original generators may still not have been recovered.}



As inertia and frequency containment reserves need to act quickly after an event has occurred, they are more crucial to the restoration of frequency.
To ensure the grid remains dependable, regulatory codes define precise requirements that these resources need to fulfil.

First, to determine the required amount of frequency containment reserves, reference incidents are considered. The frequency containment reserves are sized to be large enough to cover these incidents.
As part of this procedure, a probabilistic approach for the sizing may be used.
For this approach, sufficient reserves to cover an incident may be unavailable at most once in 20 years \cite[Article 153.2.b]{EUCOM_AS}.
Second, a minimum diversity for the reserve needs to be guaranteed.
On the one hand, resources that are part of the failure causing a reference incident need to be excluded when determining the available frequency containment reserves.
On the other hand, the contribution of any single generation unit may not exceed a total contribution of 5\% of the overall frequency containment reserves.
The stated goal of these restrictions is to ensure that a failure of any single resources will not also fail a significant portion of the frequency containment reserves that is needed in this very case to cover the failed generation of the respective resource.
Thereby, endangering the frequency restoration process \cite[Article 156.6.a-c]{EUCOM_AS}.



\subsection{Approaches to Frequency Maintenance}

Due to the increasing renewable share and the reduction of traditional synchronous generators, the inertia as well as primary and secondary frequency reserves need to be provided by alternative sources. At the same time, the fluctuations from the supply side increase \cite{ieee7738432}.
Literature shows that shifting both the generation and the ancillary service provision to \gls{DRES} is a promising solution \cite{EasyRESInertia}, \cite{ieee9508455}, \cite{ieee6268312}, \cite{ieee6466419}.

Frequency restoration using \gls{DRES} and batteries as part of the secondary frequency reserves are investigated in \cite{ieee6466419}. That approach uses optimization to reduce the probability of an outage due to a frequency instability by adding sufficient storage capacity to the volatile \gls{DRES} and loads.

Controlling the frequency is especially challenging when a grid is in islanded mode. The additional difficulty is rooted in the smaller number of synchronous generators that each need to supply a bigger relative share of the load. Therefore, a normative incident of losing the largest generation source leads to stronger frequency deviations. To overcome this the use of ultracapacitors or other fast storage systems is proposed and investigated in \cite{ieee6268312}.

Two approaches developed in the Easy-Res project achieve provision of inertia and primary frequency reserves from DRES \cite{EasyRESInertia}, \cite{ieee9508455}. 
The papers describe the communication exchange between the \gls{TSO} and one of the \glspl{DSO} for a medium voltage grid connected to this \gls{TSO} to realize the respective ancillary service. 
Both approaches follow a similar style of communication as described in the following.
The exchange can in general be split into two phases.
In the first phase, the maximum contribution to the ancillary service is estimated. 
In the second phase, a target contribution for the aggregated medium voltage grid is selected and optimally distributed to the individual resources.

For the approach to inertial response presented in \cite{EasyRESInertia}, in the first phase the \gls{TSO} informs the \gls{DSO} about the expected rate of change of frequency ($\text{ROCOF}_{max}$) in the case of the nominal incident.
The \gls{DSO} can now compute the maximum available inertial response from the distribution network using a steady state analysis for their grid $H_{ag}^{max}$.
First, the power exchanged with the external grid during normal operation $P_0^{SS}$ is determined and second, the maximum power exchange when injecting the inertial reserves $P_0^{\text{IRmax}}$ is computed. $f_n$ denotes the nominal grid frequency.
Using these variables and formula of Equation~(1) gives the maximum inertial constant $H_{ag}^{max}$ that can be offered by the system.

In the second phase the \gls{TSO} collects the received information from the distribution grids and integrates it into their respective stability analysis\footnote{Such analysis can be conducted taking into consideration the normative incidents and should be conducted for each 15 minute interval \cite{EUCOM_AS}.}.
Afterward, each \gls{DSO} receives a setting for their contribution to inertia in the form of a selected inertial constant that needs to be lower than the maximum inertial constant determined in the first step, i.e., $H_{ag}^{max} > H_{ag}^{TSO}$.
By equivalence transformations of Equation~(1), a new equation to determine the total power at the point of connection with the upstream grid can be derived (Equation~(2)).

\begin{equation}
    H_{ag}^{max} = \frac{f_n}{2} \cdot \frac{P_0^{\text{IRmax}} - P_0^{SS}}{\text{ROCOF}_{max}}
\end{equation}

\begin{equation}
    P_0^{IR} = 2\cdot H_{ag}^{TSO} \frac{1}{f_n}\text{ROCOF}_{max} + P_0^{SS}
\end{equation}

Based on this total power injection the individual power injection for each \gls{DRES} can be determined. For example, using a cost-optimal power flow computation. In the reference, line losses, converter losses, and voltage deviations where combined to determine the cost but other monetary costs can be added or substituted as well. Based on this maximum power injection and the previously established inertial constant from the \gls{TSO} each \gls{DRES} can be assigned an individual inertia constant. This inertia constant is sent to the local controller of the \gls{DRES} to control the power injection based on the actual rate of change of frequency experienced by the system.

\begin{figure}
    \centering
    \includegraphics[width=\linewidth]{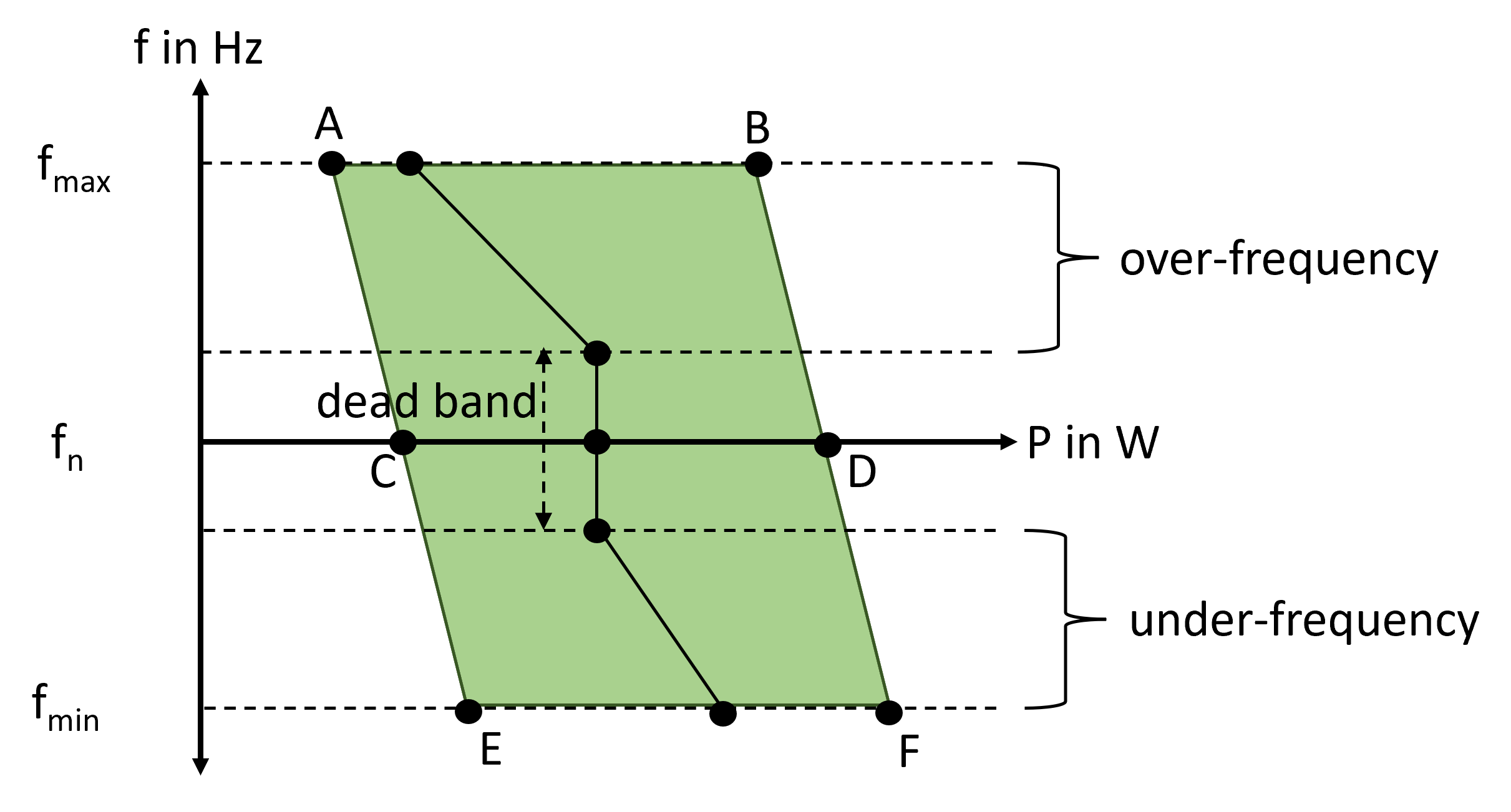}
    \caption{Feasible area for droop curves with droop curve selection.}
    \label{fig:droop_example}
\end{figure}

For the approach to provision of primary frequency reserves described in \cite{ieee9508455}, \gls{DRES} are required to operate with a headroom under normal frequency conditions.
This operational headroom is required to enable an increase of output power to cover incidents where a frequency dip occurs.
To provide the primary frequency reserves, the DRES are configured to follow a P(f)-droop curve as shown in Fig.~\ref{fig:droop_example}.
The droop curve itself can be fully defined by the dead band size, power at nominal frequency and one power-frequency pair for each area, given as ($P_{min}, f_{max}$) and ($P_{max}, f_{min}$), to compute the slopes of the respective line. 

This enables them to contribute to primary frequency reserves, as they are described above, because the \gls{DRES} are now able to fulfill the requirement of proportional response to the frequency deviations.
From a remuneration perspective this operation incurs a loss of revenue due to the decrease of \gls{PV} production in normal operation conditions that the \gls{PV} owner needs to be compensated for.
However, apart from the droop controller no additional hardware is required to be installed in the distribution networks.

With these preconditions are established, the communication steps can be explained.
In the first step, the \gls{TSO} fixes the minimum and maximum frequency parameters as well as a frequency step size. These parameters are then transmitted to the \glspl{DSO}.
Based on this information the \gls{DSO} computes a minimum and maximum power injection value for a varying set of frequencies.
These frequencies start at the minimum frequency and are repeated in regular distances with the frequency step until they reach or exceed the maximum frequency.
Two power flows need to be computed for each frequency to determine the maximum and minimum power available at the point of connection to the upstream grid.

These frequency and power range pairs creates an area. At the beginning of the second step, the \gls{TSO} can now select a desired droop curve based on a stability analysis performed for the whole grid. An example of such area is depicted in Fig~\ref{fig:droop_example}.
This area is based on computations of minimum and maximum power injections for the three frequencies $f_{max}, f_n, f_{min}$ to define its border.
A and B denote the minimum and maximum power at $f_{max}$. C and D mark the power values at $f_{n}$. Finally, the power values at $f_{min}$ are given by the point E and F.

Afterward, the selected droop curve is sent from the \gls{TSO} to the \gls{DSO}.
To achieve an optimally distributed to the individual \gls{DRES}, a series of power flows is computed to determine the power injections of the \gls{DRES} for different frequencies. 

These two approaches give tractable examples of the required information that needs to be exchanged with the \glspl{DRES} to enable contribution to frequency-related ancillary services. 
However, while the power systems side is well explored in the references, considerations the ICT side are still lacking.
The optimization approach is solely focused on grid losses.
When considering the requirement regarding diversity at a large scale in \cite{EUCOM_AS}, inclusion of path diversity or similar metrics between the connection to the outside grid and the distributed resources may give additional guarantees on the number of grid faults or communication faults that can be tolerated.
Limiting the maximum contribution of each device might enable the aggregated medium voltage grid to provide a guarantee that even when single resource failures occur the amount of reserve that is lost can be limited.
As these measures would restrict the set of feasible solutions for the optimization approach further, a potential tradeoff between operational costs and diversity needs to be considered.

From an adaptability standpoint, these two approaches are equipped with several parameters that allow them to deal with changes. The inputs sent by the TSO as part of the first communication step allow the approaches to be configured to work with larger scale modification of the grids structure. By adjusting these parameters accordingly, a TSO is able to increase or reduce the amount of inertia provided by DER in the grid. Similarly, the aggregated droop curves allows adjusting the load sharing between the grid at different frequencies by adjusting the slopes in the over- and under-frequency areas.

Reconfiguration of the grids within the medium voltage area can be handled transparently to the upstream grid as long as the aggregated behavior can remain the same. If the grid topology changes, this can be propagated to the optimization algorithm and new optimal distributions can be found for the following distribution of ancillary services to the DER.
The installation of new PVs or other \gls{DER} and consequently the integration into the provision of primary frequency reserves is enabled by adjusting the droop curves for the internal distribution to achieve the desired load sharing.

\section{Short Circuit Protection}\label{sec:ShortCircuit}


Short circuits are common faults that affect single lines. The cause for short-circuits is a low resistance connection between two lines or a line and the ground. For example, a bird's wings touching two lines, wet vegetation interfering with the power line, or a human flying a kite can cause this. This short cut in the electrical circuit causes a large amount of current to flow immediately towards the fault. This current can cause damage to equipment and endanger human lives.
To prevent this current lasting long enough to cause damages, the power system is equipped with a protection system that is disconnecting the faulty line from any power supply. \nhl{Therefore, the proper operation of the protection system ensures safety.}

This is achieved by breakers, current-sensitive devices that are connected between two electrical conductors and can disconnect this line if the current exceeds some limit. The process of disconnecting is also called tripping. Due to the traditional top-down operation mode of the power system, the protection system is built in a hierarchical manner. That means, breakers that trip at lower current are located closer to the customers and breakers with higher current limits are located closer to the external grid connection and to large-scale generation sources in the transmission network.

The increasing inclusion of \Glspl{DER} breaks this structure. 
Due to large distributed generation, the current flow may now invert to start at customer sites. Consequently, changing current magnitude all over the grid.
Furthermore, if large synchronous generators are decommissioned in the upstream grid, sufficient fault current may not be available from an external source only and the participation of \gls{DER} in the provision of this current is required. \nhl{Due to this change, the protection system faces several new challenges, which are explained in the following section. Afterward, how the power system protection can remain resilient despite the inclusion of} \gls{DER} \nhl{is highlighted.}

\begin{figure}
    \centering
    \includegraphics[width=\linewidth]{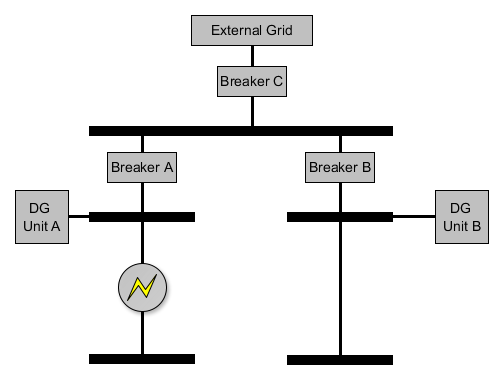}
    \caption{Simple two-feeder system with distributed generation.}
    \label{fig:twoFeeders}
\end{figure}

The baseline for this explanation is the simplified two feeder power system shown in Fig.~\ref{fig:twoFeeders}. A fault occurs in the left feeder while two energy sources (DER Unit A and B) inject power. Breakers A and B share the same sensitivity while Breaker C can sustain a higher current. In an operation mode without \gls{DER}, the external grid would provide sufficient current to trip exactly Breaker A while keeping the current low enough for Breaker B to not trip. This means the tripping only occurs in the faulty feeder and the healthy feeder remains unaffected. 

\subsection{Issues Caused by Distributed Generation}

For this section, the relevant aspect of the service provided by the power system is its ability to continuously supply the load with power in a safe manner. This service can then be degraded if the service is no longer safe due to excess currents or if part of the load is disconnected.
\nhl{Going back to the resilience state space, one can then interpret an unsafe, excess current flow as being unacceptable for the service provided by the power system.
The disconnection of the breakers is then part of remediation that is triggered after a high amount of current is detected. The resulting service provided by the power system leaves the lines behind the breaker without power. Therefore, the service level is still impaired but at least no longer dangerous to the environment of the power system.} The inclusion of \gls{DER} leads to several problems related to the detection and remediation discussed in the following.

First, consider the case were indeed the protection system operates properly and trips breaker A to disconnect the left feeder from the external power supply.
If DER unit A is still injecting current during this period, the line will remain energized.
\nhl{That means the remediation action was engaged, however, the service provided by the power system is still unacceptable,} because there may still be a high current flowing towards the fault.
This exposes the maintenance personnel, which is sent to investigate and recover the system by clearing the cause of the short-circuit, to the risk of electric shock.
If only insufficient monitoring information is available, the state of the \gls{DER} and therefore, the state of the line may be unknown. This is despite the open breaker clearly indicating the line is disconnected from the power supply.

\begin{figure}
    \centering
    \includegraphics[width=\linewidth]{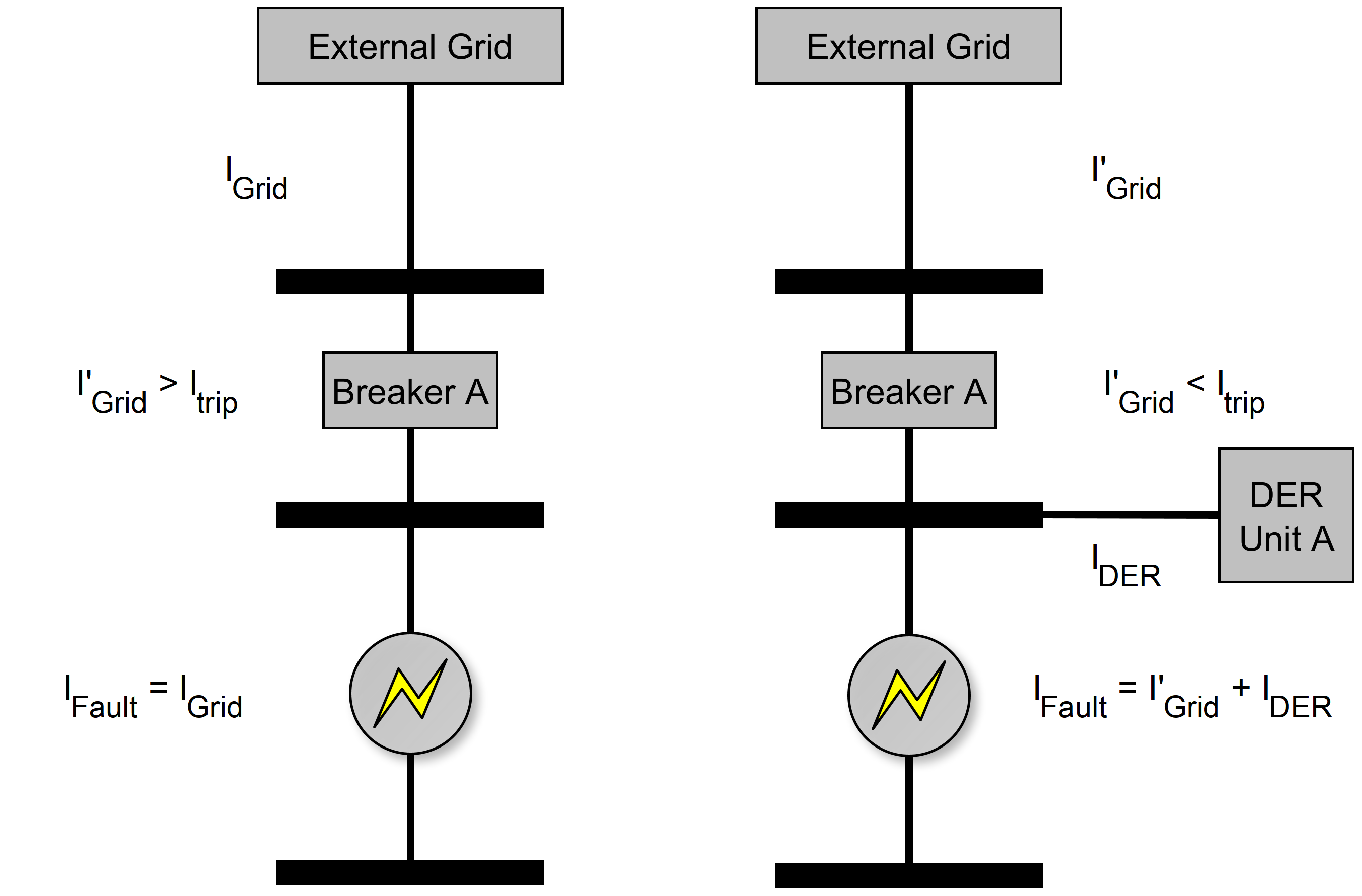}
    \caption{Protection blinding in a feeder with a DER system.}
    \label{fig:ProtBlinding}
\end{figure}

The second problem is called protection blinding and is shown in Fig.~\ref{fig:ProtBlinding} as a comparison between scenarios with and without distributed generation.
Protection blinding describes a situation in which a breaker that should trip does not trip due to insufficient current flowing.
\nhl{This fails the detection step of the protection systems operation and no remediation of the service level is performed.}
To understand this issue a closer look at the relation of the currents is required.
The fault will cause a certain amount of current to flow.
The exact amount depends on the type and location of the fault.
Let this current be denoted as $\text{I}_{\text{fault}}$.
The current required to trip the breaker is called $\text{I}_{\text{trip}}$.
In the case without \gls{DER}, a current $\text{I}_{\text{grid}}$ is flowing from the external grid towards the fault while traversing Breaker A.
In this scenario, all the required current from the fault is supplied by the upstream grid.
That means $\text{I}_{\text{fault}} = \text{I}_{\text{grid}}$. 
Therefore, the breaker will disconnect, because it is configured such that $\text{I}_{\text{grid}}$ is greater than $\text{I}_{\text{trip}}$.

With \gls{DER} in the same feeder as the fault, the scenario changes. 
The DER unit is providing an additional current $\text{I}_{\text{DER}}$.
This reduces the current drawn from the external grid. The exact amount of current depends on many inherent characteristics of the power system such as line impedances, location of the fault, and type of \glspl{DER}, which is beyond the scope of this explanation. However, qualitatively speaking it holds now that $\text{I}'_{\text{grid}} < \text{I}_{\text{fault}}$. 
If the setting of the breaker still is configured for the old $\text{I}_{\text{grid}}$, the new current may be smaller than the tripping current $\text{I}_{\text{trip}}$.
In turn, the current is insufficient to cause the breaker to trip.
As a result, the tripping is delayed until other mechanisms are triggered (e.g., detecting a larger current over a longer period) or may never happen preventing proper disconnection of the line. 

\begin{figure}
    \centering
    \includegraphics[width=0.9\linewidth]{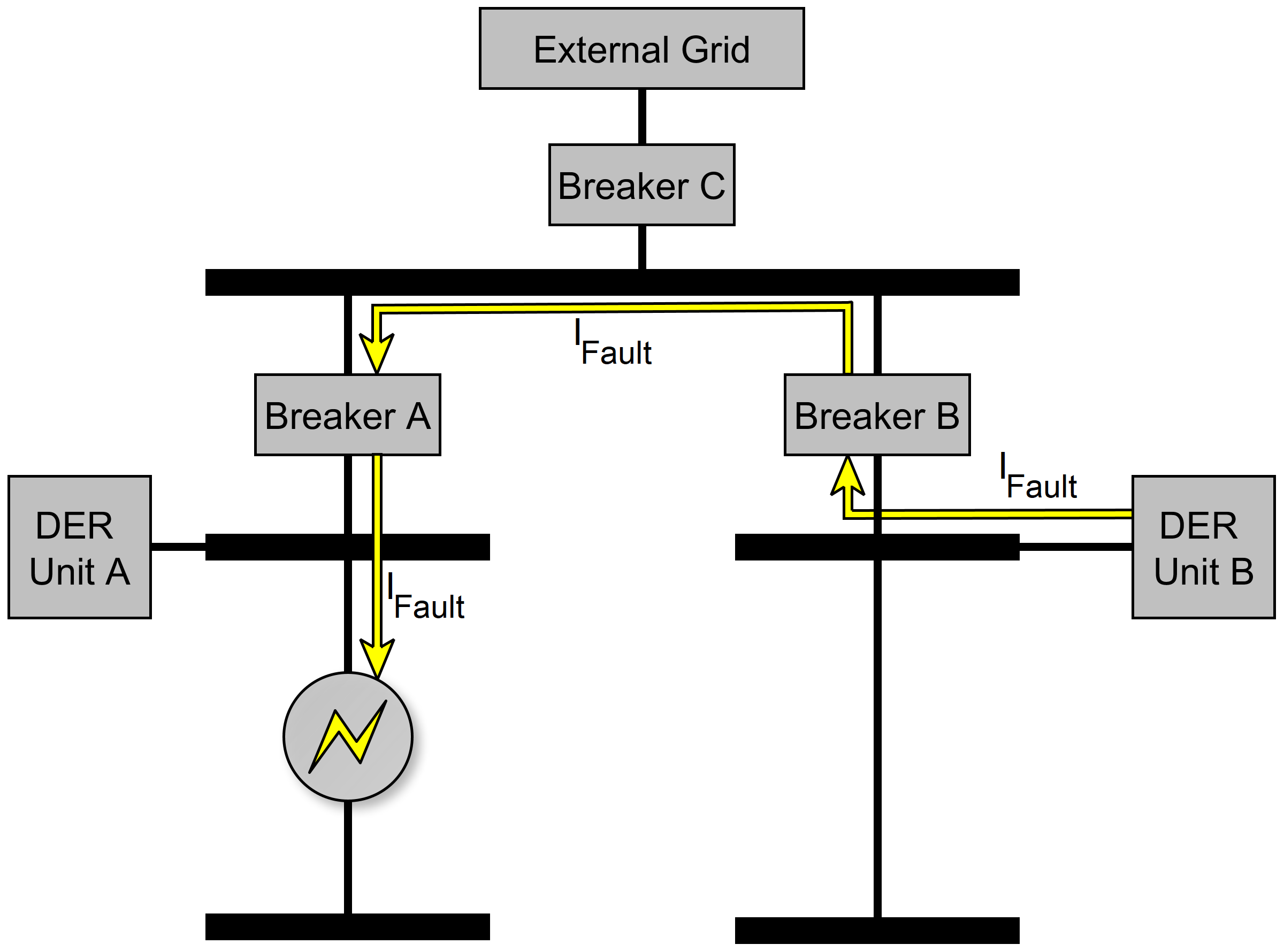}
    \caption{Sympathetic tripping in a two-feeder system.}
    \label{fig:SympTripping}
\end{figure}

The third issue caused by \gls{DER} has two additional pre-conditions.
First, no sufficient fault current is available from the upstream grid and second, the fault current is instead to be sourced from local \gls{DER} in an adjacent feeder.
In this case, the current provided by \gls{DER} in the feeder without a fault can exceed the tripping-limit of the breaker in the fault-free feeder. Such current flow is depicted also in Fig.~\ref{fig:SympTripping}.
This may cause simultaneous tripping of two breakers.
One cause for this may be a shortcoming of the detection via breakers. 
In other protection schemes, that rely also on a monitoring of the system state to detect faults, second cause for this problem may the a failure to control the current injections from the healthy feeder to remediate the issue.
In both cases, the wrongly performed remediation has a two-fold impact.
One the one hand, this doubles the effort to locate and fix the faults\nhl{, i.e., recovery of the system state is hindered.} On the other hand, the number of loads that are disconnected from the power system increases, \nhl{which leads to a worse service level during the interruption.}
In the example grid, one of the assumptions stated earlier is that both feeders share the same current limit for their respective breakers.
Without injection from the external grid or a third feeder, it is therefore not possible to avoid the protection blinding issues in the example, because if DER Unit B is providing sufficient current to trip Breaker A, this current will also trip Breaker B.
Even if such supply might be available, it is still required to coordinate the currents supplied by \glspl{DER} and the current limits for the breakers to avoid a \gls{DER} injecting a too large current and tripping the breaker of its feeder.

\subsection{Approaches to Adaptive Protection Schemes}

The core of these three issues is a mismatch in the injected current by the \gls{DER} or the upstream grid, and the reaction of the protection system to these. \nhl{A resilient smart grid needs to coordinate these devices in a way to ensure the dependability of the protection system.} This is known as adaptive protection.



For the adaptive protection schemes outlined here, two general paradigms can be adopted based on where the decision is made to trip a protection device in the grid: distributed schemes and centralized schemes.
In distributed schemes, multiple relays coordinate their operation based on their local view of the system without a centralized controller. The authors of \cite{ieee5871328} showed one such approach that pre-configures relays with multiple settings to be followed during different grid configurations (islanded versus grid-connected mode) or when certain generators are connected or disconnected.
\nhl{This leads to a resilient protection system, as despite environmental changes, e.g., the islanding of the grid, the protection system can remain dependable.}
\nhl{Going one step further, dependability of the protection system is also ensured between the occurrence of a fault and its clearing}, by using the time delay for clearing a fault observed by all relays upstream of the fault. This makes it possible to determine which generation units are disconnected during the fault and consequently update the relay settings until the fault can be cleared.

A second approach shown in \cite{ieee6750126} is based on a multi-layer scheme. A lower layer is tasked with making the decision to trip a breaker. A second higher layer is tasked with optimizing the settings for this response. \nhl{Similar to the previous approach, this makes the protection system resilient.} 
The added value of \cite{ieee6750126} is that this operation of the protection system is possible even if part of the communication system has failed.
In case of a communication failure with the optimization layer, the current configuration is kept until communication is restored. 
\nhl{Even though this ensures that the system configuration remains consistent even if communication is unavailable,}
this proposal still requires the communication system to be dependable in the end.
As soon as a link failure is present and a re-configuration is issued the state of the relay connected via the failed link is uncertain to the optimization layer.
Locking out the system can lead to problems in scenarios that require re-configuration.

Which actions to take when re-configuration of the protection system is only partially available is still an open research question.
Re-configuration of the other breaker settings might run into mis-configuration if a wrong assumption is made about the state of the unreachable relay.
Sticking to the old settings may leave the protection system in an invalid state if the power flow in the power system changes.

A multi-agent based approach is presented in \cite{ieee5523996}. 
Agents are created for every relay and every \gls{DER}. Based on an information exchange between the \gls{DER} agents and relay agents, relay configurations are determined. The correct profile is selected depending on which \gls{DER} units are connected. Additionally, as part of this communication the status of the \glspl{DER} is disseminated.
This again enables the proper configuration of breaker limits \nhl{and therefore the resilience of the protection system.}
\nhl{The dependability of the protection system is increased further by including a backup mechanism for in case the reaction of a relay is absent due to protection blinding.}
This is achieved by an additional exchange between relay agents to correctly identify the fault location relative to the breakers and send backup trip signals if a breaker is unable to open in the present circumstances. The same communication mechanism enables to lock out a relay that should not trip if another relay is determined to be better suited to isolate the fault. Therefore, this mitigates the issues of sympathetic tripping. As there is communication during the fault clearing, real-time requirements apply for the communication delays which \cite{ieee5523996} does not comment on.

In centralized schemes, a central coordination unit is collecting measurements in real-time and sends actuation signals to breakers in the system. This requires the communication system to fulfill strict bandwidth and latency guarantees. This highlights that both, distributed and centralized schemes, need dependable ICT with high quality-of-service guarantees.

Regarding centralized schemes, promising results regarding fault isolation are achieved by \cite{ieee1256357} given that sufficient measures for preparing the system are taken. Then, the approach can effectively deal with the above mentioned issues in protection coordination stemming from \glspl{DER}.
First, additional grid hardware is required in the form of phasor measurement units for measuring and synchronizing the current measurements on all three phases, remote controlled breakers for individual parts of the system, as well as directional current measurements for all these breakers.
Furthermore, sufficient \gls{ICT} resources to communicate and process this information are required at a central unit.
Crucial to the operation of the system is the operation of a central relay that sends disconnection signals to the breakers in the grid.
Based on pre-computed information in an offline manner, the scheme is able to identify the location of faults using the thevenin-equivalent impedances\footnote{The thevenin impedance for a fault can be computed by replacing voltage sources in the system with short-circuits and current sources with open circuits and then applying the laws for serial and parallel circuits regarding impedances.} and a three phase model of the power system to compute power flows for the short circuit analysis.
If the injections of the \glspl{DER} are increased, it can be determined whether a fault has occurred or not using this information. Furthermore, due to the different impedances of the lines leading from each DRES to the fault, the current injections for each DRES are characteristic for the location of the fault. This relation is previously estimated based on the thevenin-equivalent impedances. Therefore, the fault location can be determined and signals can be sent to the respective closest breakers to isolate this fault.
Due to the unique identification of the fault location and the corresponding breakers, this approach can mitigate the above mentioned issues of protection blinding and sympathetic tripping.

One can observe several problems that occur for the resilient operation of this centralized scheme.
Even though a lot of information is generated in the offline phase, only a single centralized decider is responsible for the grid area.
Failure of this device leads to a failure of the overall protection where it is difficult to recover from without recovering the centralized unit.
Furthermore, a large stream of information needs to be analyzed continuously during the operation of this scheme. 
The possibility of data corruption on the communication links as well as of intentional attacks of wrong data injection require that additional measures are taken to secure this communication. However, strict communication requirements need to be fulfilled for these schemes to operate in real time.
On the other hand, one advantage of this approach is the possibility to still isolate the faulty part of the grid when single breakers are not responding to the disconnection signal. This is achieved by continuing to measure the current provided by the \gls{DER}, noticing that the fault persists and then escalating to disconnection of a larger part of the grid until the fault is cleared as described in \cite{ieee1256357}.

The approaches of \cite{ieee8288855} and \cite{ieee7169618} do not use an ICT system but instead rely only on distributed decision-making.
However, this does not provide the ability to integrate measurements, forecasts, and topological changes into the power system, which in the worst case may be to switch to an islanded operation mode for the microgrid. 
As these are desirable properties in grids with high renewable penetration, an \gls{ICT} system becomes necessary to handle these tasks.
Nevertheless, care must be taken to not overstate the performance of the \gls{ICT} system. 
The approaches in \cite{ieee5523996}, \cite{ieee1256357}, \cite{en11020308}, \cite{ieee9600981}, and \cite{ieee8556478} make the unrealistic assumption of a perfectly working \gls{ICT} system.
This assumption oversimplifies the situation, as in an interconnected system failure is not an exceptional case but rather the norm.
Especially, the approaches of \cite{ieee5523996} and \cite{ieee1256357} that require communication during the short circuit event, need to be extended with strategies that explain the reaction of device when faced with a temporary disruption of the communication service.
Of the presented approaches only \cite{ieee6750126} adds considerations of behavior during communication faults. 

\section{Recovery from Blackout}\label{sec:Blackout}

\begin{figure}
    \centering
    \includegraphics[width=0.75\linewidth]{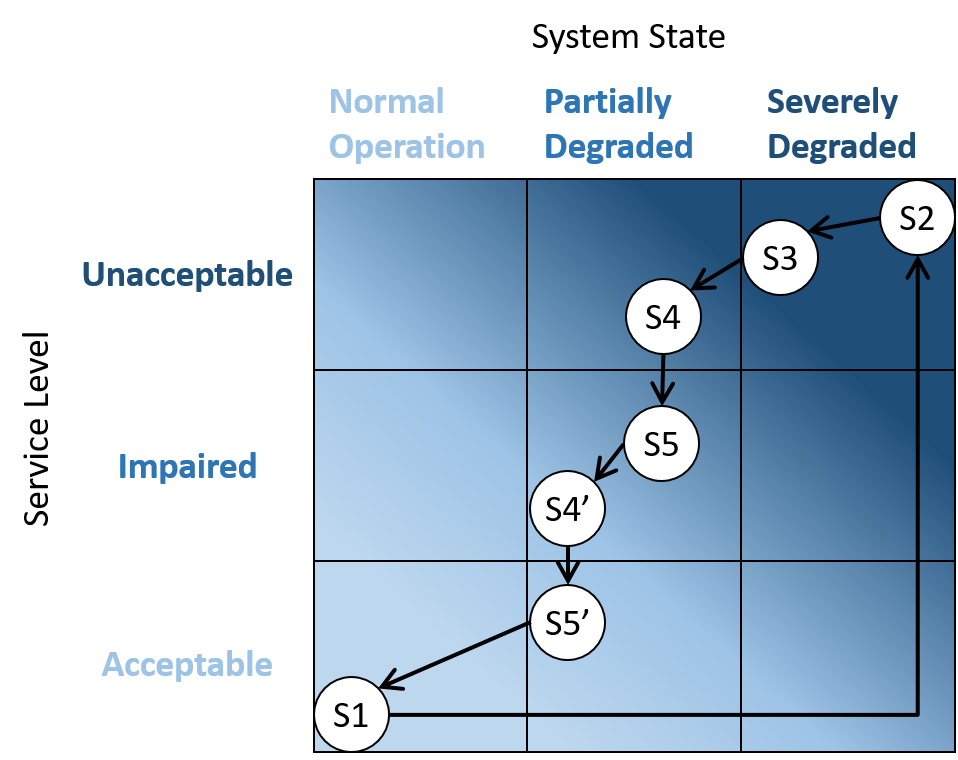}
    \caption{System state during the black start process.}
    \label{fig:BlackoutOverTime}
\end{figure}


A blackout refers to the total collapse of the power system.
Such a collapse may be caused by the failure of the frequency reserves to cover a large frequency drop, i.e., more generation power is lost than considered in the reference incidents.
This may lead to a loss of synchronization between the generation sources. 
The resulting cascading series of events eventually leads to the total collapse of the system.
The root cause for the failure such chains of events could for example be natural disasters that cause severe damage to the transmission and distribution lines as well as the generation equipment.

\nhl{Speaking in terms of resilience, a blackout is the consequence of failed remediation of a power system. Therefore, these scenarios are more likely to occur if the inertia and reserves of frequency response decrease.}

During a blackout, loads are no longer supplied with electrical energy. 
Failing to supply certain types of consumers, such as hospital or other infrastructures, may have more severe consequences than for other types of load. 
For these critical load to remain operational in face of a blackout, emergency generators or batteries need to be present to cover the local consumption.
However, these emergency sources can only operate for a limited amount of time determined by the energy reserve, e.g., the energy stored in a battery.

The lack of energy also affects the whole \gls{ICT} infrastructure that is coordinating the smart grid. 
Therefore, monitoring and re-configuration tasks may be impossible, as the required resources are currently unavailable due to a lack of power.
Nevertheless, in the setting with predominantly distributed generation, it is necessary to coordinate these resources to operate and consequently recover the grid.
In summary, the \gls{ICT} system relies on the power system for energy supply while at the same time recovery of the power system requires configuration or control via the \gls{ICT} system. 
The recovery from a blackout is also referred to as black start.
As a result, any successful black start procedure needs to overcome this cyclic dependency to recover the system state.

Traditionally, black start is performed using a top-down approach to coordinate a few large generators in the transmission network \cite{ieee8939530}.
\nhl{The change in the generation scheme from centralized, synchronous generators to} \glspl{DER} \nhl{may lead to scenarios where sufficient generation for grid recovery is not available in the high voltage grid. This may be the case if synchronous generators are sparsely distributed in the grid and damage to transmission lines leaves some areas without a connection to a synchronous generator.
Therefore, a top-down paradigm may no longer be feasible.
Recoverability is part of a dependable system even from the worst-case scenario described here. Therefore, a resilient smart grid needs to maintain this property using an alternative to the traditional schemes in a bottom-up approach heavily relying on} \glspl{DER}.

However, not every type of \gls{DER} is capable of re-starting its local grid.
To achieve this type of local supply, the \gls{DER} needs to establish both the correct voltage and frequency for the grid on its own. According to \cite{ieee6200347}, this partitions the set of \gls{DER} into the three types grid-forming, grid-supporting and grid-feeding.
A device that is able to act as a voltage-source and establish voltage and frequency without another reference is referred to as a grid-former or as having grid-forming capabilities.
A grid-supporting device is only capable of regulating these parameters once they are established, i.e., it requires a voltage and frequency it can measure and correct if necessary.
Finally, grid-feeding \gls{DER} act as current-sources that means they inject a certain amount of power into the grid without measuring and reacting to fluctuations. The injected power can be determined by what is available from the primary source, e.g., in the maximum power point tracking control of a PV, or following a target set by a higher level controller, e.g., a battery charging and discharging according to a pre-defined schedule.
The availability of grid-formers is an important pre-condition for any approach to distributed black start.
Literature shows this role may be fulfilled by small synchronous machines, such as combined heat and power plants \cite{chp_blackstart}, or \gls{DRES} installations using a converter with these capabilities included in its control \cite{ieee8810684}.

The bottom-up recovery of the power system using \gls{DER} happens in multiple stages.
Figure \ref{fig:BlackoutOverTime} shows the different stages and the development of served load during the recovery procedure.
The individual phases are adopted from \cite{blackstart}, which is detailed at the end of this section.
The service level provided by the smart grid system in this case is measured by the amount of served (critical) load. If the critical load can only be partially supplied, the service level is unacceptable. If the critical load can be supplied, but not all other loads then the service level is impaired. If all of the load, critical and non-critical, can be supplied the service level is deemed acceptable.
On the other hand, the operational state is determined by the state of the available power sources.

Before a blackout occurs, a system is in the normal operation state, in which it is able to supply the full load (S1).
Immediately after a system collapse, no load can be served. That means the service level and system state are maximally degraded (S2).
Next, the individual recovery processes of the \gls{DER} units with grid-forming capabilities restart.
Therefore, the operational state of the system is improved. These resources offer the capability to partially supply the local load and therefore improve the service level (S3).
In this context, these independently supplied grid areas that have smart controls via an \gls{ICT} system for black start and other ancillary services available are referred to as microgrids.

Once the grid-forming devices have established frequency and voltage for the microgrid, additional grid-following resources can be re-connected to stabilize these microgrids further. 
This again improves the operational state and the service level (S4).
Depending on the type of non-black start capable resource, different preconditions for the re-connection need to be ensured. 
Some resources may need a stable voltage and frequency profile as they are configured to disconnect at a certain threshold.
Other resources may require auxiliary loads as part of their start-up procedure. Afterward, they can supply the required power for these loads themselves, e.g., power for a water pump in thermal generation units or the required power supply for local control and communication systems \cite{ieee7438896}.


The next step is to re-connect the individual microgrids with each other.
This requires synchronizing the frequency of both areas, similar to the re-connection of an islanded microgrid to the main grid explained in \cite{ieee7223046}. To achieve a smooth re-connection the frequency of the two grid areas need to match closely. According to the reference, this is the case if the frequency magnitude is the same and the relative shift of the two sinusoidal oscillations is less than 0.2 radians.
To achieve this, the \gls{DER} need to be informed about the differences in the frequency. With this information, the \gls{DER} can adjust the local frequency accordingly. It should be noted that the reference \cite{ieee7223046} assumed a sampling rate of 10 kHz for the operation of the \gls{DER} when adjusting the frequency signals. Although the exchanged information are only two measurements of differences in the frequency for each instance, this high data rate and the requirement for close synchronization make a reliable ICT system necessary.
This synchronization process does not improve the system state in terms of operational generators.
Instead, the grid stabilizes further as excess generation in one area can cover for excess load in other areas.
Therefore, the operational state stays the same, but the service level increases (S5).

In some case the grid are that is re-connected with the existing microgrid has no grid forming resources available, i.e., it is currently un-supplied. In these cases, the re-connection step of the grid-following resources might be repeated (S4'). The synchronization with neighboring microgrids continues until a maximum extent is reached (S5'). Numerous conditions influence where this stopping point lies, e.g., damage to transmission or distribution lines, damage to equipment, requirements for synchronization across the transmission system level. Here, it is assumed that the energy provided by the microgrids is sufficient to restore the grid following loads in the transmission system and perform full recovery to S1, similar to the recovery of a large-scale plant described in \cite{ieee7438896}. 

Afterward, in the re-connected microgrids, it is possible to use supply from one microgrid to supply the loads in the other. As a result, more load can be supplied with electricity. 
This might make it so that further communication resources become available to repeat this process with other microgrids that were previously unable to communicate with each other.
After some of these recovery steps are performed, the critical load in the system can be fully re-supplied by the \gls{DER}. For the determination of service level, this is selected as the threshold to move from unacceptable service to degraded service.
To restore an acceptable service, this process iterates further with stepwise increases until the whole grid is recovered. 

\subsection{Approaches to Black Start}

Several approaches exist that deal with the microgrid formation and synchronization in the black start scenario.

The approach of \cite{en11123394} shows iterative restoration of power supply to a grid.
First, battery storages enable restarting wind turbines. 
These are then used as the second restoration step.
Afterward, the grid is stabilized by switching on supporting sources such as thermal generation units.
Note that these steps align with the microgrid formation process described above where first the black start capable units, wind farms equipped with batteries in this case, are started before supporting sources join in to stabilize the grid.

Another question that arises in this context is the appropriate sizing of microgrids. This is the topic presented in \cite{en10070948}.
That paper investigates how to determine the best allocation of independently recoverable areas of a power system. 
As is shown in \cite{en11010001}, this will enable faster recovery after a blackout as independent areas can be restored in parallel. Furthermore, the authors of \cite{en11010001} investigate the next step of the recovery procedure once parallel recovery is done. Doing so requires determining how to best interconnect these microgrids.
It is also noteworthy that multiple different grid topologies can be assumed for the sub-grid to be recovered.

The approach in \cite{en9050372} proceeds overall similar, but assumes a hierarchy with a three-phase microgrid on the top that has several single-phase microgrids connected to it.
Such a topology requires different control strategies for the standalone operation mode of the single-phase and the three-phase microgrids and a sophisticated re-connection strategy to deal with the increased complexity in the interaction between phases.


The way to recover the grid proposed in the Blackstart project is investigated further in this tutorial \cite{blackstart}.
That approach makes the following assumptions:
First, the upstream grid is assumed to be failed after a disaster and remains failed throughout the recovery.
Second, there are multiple \gls{DER} with grid-forming capabilities located in the medium voltage network.
Third, a larger amount of grid-supporting resources able to supply the demand are available, but cannot be connected until the grid is stabilized.
Fourth, in several places switches are located in the power system to control the connection between parts of the grid.
Under these preconditions, an iterative recovery process between power and \gls{ICT} system is built based on a multi-agent system.
At a high level, first localized supply of distributed generators is used to form microgrids and re-established supply of local loads.
By doing so, parts of the communication network can be restored as well.
Afterward, neighboring parts of the grid are able to communicate, reconnect and increase the supplied load.

In more detail, an agent is created for every grid-forming generation unit, load, and every grid area that can be separated using switches.
An agent responsible for a grid area acts as an aggregation point of the load and generator agents in this area.
First, the agents responsible for the areas collect and exchange information about their loads and generation available.
Next, these agents derive a target schedule for the power production of generators and disseminate this to the load and generator agents. Note that this schedule includes also the capability to supply load not only within the local area but also in the area of other area-responsible agents.
Afterward, the load and generator agents coordinate to fulfill the target schedule as good as possible. 
The result of this coordination is returned to the area-responsible agent.
If this changes the information known to the agent, another exchange among the agents responsible for the sub-areas is started.

The approach is evaluated in simulation when varying two parameters related to the \gls{ICT} system: the available backup storage for communication node and the range covered by communication nodes. The available backup storage for the communication nodes is distributed randomly.
The experimental results show that more energy in the emergency batteries increases the total restored load.
The same holds for an increase in the cell size.
Different locations of the emergency batteries lead to considerable variation in the restored load. 
A final note from that approach is that cell and battery size can compensate for each other.
For example, restoration of at least 50\% of the load is possible if availability of battery storages is 90\% for 2km cell size, 50\% for 6km cell size or 10\% for 10km cell size.
Almost complete restoration of the load, i.e., more than 80\% restoration, is only reliably possible with 90\% availability of the battery. For a cell size of 2km, this is possible in about 80\% of the simulation runs. Furthermore, this is possible with a cell size of 6km or larger in all simulation runs.
The study shows that with little battery reserves and high cell size similar amount of load can be restored as with high battery reserves and low cell sizes.

The authors of \cite{blackstart} highlight further research questions related to their approach.
As the location of battery reserves matters, restoring energy supply to certain, critical nodes in the system first can improve the convergence time and restored load for this approach.
Further considerations are concerned with larger than anticipated forecast errors for the available generation of the renewable sources and the demand of loads. Additionally, depending on the cause for the blackout other secondary challenges may exist in the power system after the collapse, e.g., uncleared short-circuits. If a grid-section with an undetected short-circuit is connected to the other grids, the resulting instability due to the large fault current may be enough to destabilize the grid causing the black start to fail. Consequently, restored areas may return to being unsupplied.


For blackout scenarios, avoiding single points of failure is crucial. 
If a central controller is required for the recovery procedure and this unit is affected by the power outage, the respective grid area cannot be recovered until this controller has been replaced.
However, the hierarchical control proposed in \cite{en11010001, en9050372} requires such a controller located at the distribution system level.
\cite{en10070948} and \cite{en11123394} make no explicit reference to the required communication structure. Nevertheless, determining the grid sections \cite{en10070948} requires computation-intensive optimization that indicates a centralized scheme, whereas for the iterative approach described in \cite{en11123394} a decentralized approach may be used.
Finally, the multi-agent approach of \cite{blackstart} can proceed despite partial failure or unavailability of agents even though only a smaller portion of the load may be restored.

\section{Conclusion}\label{sec:Conclusion}



In conclusion, enabling a power system to be resilient, i.e., ensuring its dependability despite a replacement of synchronous generators with renewable sources, requires to ensure sufficient provision of ancillary services.
For three critical ancillary services, frequency regulation, short-circuit protection and black start, this paper shows that the coordination of power system resources via an ICT system is able to continue their provision despite current and future changes.

As decommissioning of synchronous generators decreases inertia and primary frequency reserves, frequency regulation is considered first.
These reserves are critical for the remediation of the power system, which is proven by mapping the behavior of the system in response to a large generator failing to the resilience state space.
Investigating approaches from literature shows that maintaining the dependability of a power system in this regard requires an ICT system to aggregate the available resources, determine the optimal setting per \gls{DRES} and distribute the required configuration for stable grid operation.
This is validated with approaches for inertia and frequency reserve from the EASY-RES project \cite{EasyRESInertia}, \cite{ieee9508455}.

Next, the protection system is investigated. Here, distributed generation is the root cause of protection blinding, sympathetic tripping and isolated power lines still being energized.
From a resilience perspective, these problems are different types of incorrect operation of the detection and remediation mechanisms for grid faults.
To maintain safety in a power grid, it is deemed necessary in literature to either update the configuration of protection devices in the system or directly send actuating signals to them in response to a fault.
The resulting adaptive protection schemes require an ICT system that is highly available to correctly distribute configurations and that offers high performance guarantees regarding bandwidth and latency for time-sensitive communication in response to a fault.

Finally, recovery after a blackout is investigated.
Without a proper power supply, only a portion of generation resources in a power system, the so-called grid-forming resources, and only ICT resources equipped with batteries are operational.
An analysis of a multi-agent based approach to coordinate the power supply by grid-forming resources and the interconnection of recovered areas showed the requirements for an ICT system to enable this process.
To achieve a load restoration of over 80\%, at least 90\% of communication nodes need to be equipped with such batteries and these communication nodes need to have a communication range greater than 6km.
This shows that availability and performance of an ICT system are important factors for the success of a black start procedure. At the same time, such procedures need to deal with unavailable power and ICT nodes during the system recovery.

As is shown in each of these cases, maintaining dependability of a power system is only possible with an ICT system that is also be dependable as it needs to be available, fault-tolerant, of high performance and able to adapt to changing system conditions.
In other words, a resilient power system despite the increasing number of renewable sources requires a resilient ICT system.

\section*{Acknowledgement}

The work that lead to this paper was supported in part by the European Union’s Horizon 2020 research and innovation programme under grant agreement No. 957845: ``Community-empowered Sustainable Multi-Vector Energy Islands - RENergetic",
by the Deutsche Forschungsgemeinschaft (DFG, German Research Foundation) project number 360475113 as part of the priority program DFG SPP 1984 -Hybrid and Multimodal Energy Systems: System theory methods for the transformation and operation of complex networks, as well as by the 2022 Reykjavík Summer School on Secure and Reliable Distributed Systems sponsored by the german academic exchange service (DAAD).

\printglossary[type=\acronymtype]

\bibliographystyle{IEEEtran}
\bibliography{IEEEabrv,bibliography.bib}



\end{document}